\documentclass[aps,twocolumn,
                amsmath,amssymb,%
                pre,showpacs]{revtex4}

\usepackage[dvips]{graphicx}
\usepackage{epsfig}

\begin{document}

\title{Paradoxes of Subdiffusive Infiltration in Disordered Systems}
\author{Nickolay Korabel and Eli Barkai}
\affiliation{Physics Department, Bar-Ilan University, Ramat-Gan 52900, Israel}

\date{\today}

\begin{abstract}
Infiltration of diffusing particles from one material to another where the diffusion mechanism is either normal or anomalous is a widely observed phenomena. When the diffusion is anomalous we find interesting behaviors: diffusion may lead to an averaged net drift $\left< x \right>$ from one material to another even if {\em all} particles eventually flow in the opposite direction, or may lead to a flow without drift. Starting with an underlying continuous time random walk model we solve diffusion equations describing this problem. Similar drift against flow is found in the quenched trap model. We argue that such a behavior is a general feature of diffusion in disordered systems. 
\end{abstract}


\maketitle

Infiltration of diffusing particles from one material to another is a widely investigated process in many fields of physics. In recent years much focus was diverted to the problem when the diffusion in one or in both materials is anomalous, namely $\left< x^2 \right> \sim t^{\alpha}$ with $\alpha \ne 1$ \cite{BG,MK}. Among many examples where this behavior is important are infiltration of water into porous soil \cite{Abd}, contaminant diffusion \cite{Kirchner}, moisture ingress in zeolites \cite{Azevedo} or in fired clay ceramics \cite{Kuntz}, diffusion of sugar through a membrane in a gel solvent \cite{Kos}, and polymer translocation through a membrane pore \cite{KK}. Infiltration is also important in biologically motivated experiments. For example proteins diffusion is anomalous in the cell and normal in the exterior, compartments on membranes indicate that diffusion of proteins is taking place between different regions with varied diffusion mechanisms \cite{Jacobson}, morphogens are subdiffusing in extracellular environment where the diffusive properties changes abruptly in space \cite{Morph}.

Consider unbiased diffusion in one-dimension where one type of diffusion takes place in $x<0$ and another in $x>0$. The infiltration of particles from one material to another may lead to an averaged net drift $\left< x \right>$. We show that for subdiffusion the flow of particles may be in the opposite direction of the drift. Even more surprisingly, we find situations when asymptotically {\em all} the particles are in one sample but the average drift $\left< x \right>$ is oppositely directed. This is a paradoxical behavior in the following sense: let $P(x,t)$ be the normalized probability density function (PDF) of finding a particle at time $t$ in $(x,x+dx)$, $\int_{-\infty}^{\infty} P(x,t) dx =1$. One can argue rather generally that if $\lim_{t \rightarrow \infty} \int_{0}^{\infty} P(x,t) dx =1$, i.e. all particles are in $x>0$, then $\lim_{t \rightarrow \infty} \int_{-\infty}^{0} P(x,t) dx =0$ which implies that $P(x,t)=0$ for $x<0$ (since $P(x,t) \ge 0$) and hence the drift should be positively directed $\lim_{t \rightarrow \infty} \left< x \right> = \int_{-\infty}^{\infty} x P(x,t) dx =\int_{0}^{\infty} x P(x,t) dx>0$. Paradoxically in some cases of subdiffusion we find the opposite behavior $\lim_{t \rightarrow \infty} \left< x \right> < 0$ (even though all particles are eventually in $x>0$). Similarly, in some cases $\lim_{t \rightarrow \infty} \left< x \right> = 0$, but in the long time limit all particles accumulate in one sample, for example in $x>0$, $\lim_{t \rightarrow \infty} \int_{0}^{\infty} P(x,t) dx=1$. The solution of these paradoxes is given in this Letter, as well as a derivation of $\left< x \right>$ and $\int_{0}^{\infty} P(x,t) dx$. 

{\em Model 1: Fractional diffusion equations} \cite{MK,FD}.---Consider the semi-infinite region $x<0$ which has subdiffusive dynamics with exponent $0 < \alpha^{-} \le 1$ and diffusion constant $K^{-}$ whose units are $\text{mt}^2/\text{sec}^{\alpha^{-}}$. Similarly for the domain $x>0$ the exponent $0 < \alpha^{+} \le 1$ and $K^{+}$ $\left[ \text{mt}^2 /\text{sec}^{\alpha^{+}} \right]$ govern the dynamics. Subdiffusive processes are described by fractional diffusion equations \cite{note} (see also \cite{Morph})
\begin{eqnarray}
\label{anomaleq}
\frac{\partial P(x,t)}{\partial t} = \; _{0}D_{t}^{1-\alpha^{-}} K^{-} \frac{\partial^2}{\partial x^2}P(x,t), \quad x<0, \nonumber \\
\frac{\partial P(x,t)}{\partial t} = \; _{0}D_{t}^{1-\alpha^{+}} K^{+} \frac{\partial^2}{\partial x^2}P(x,t), \quad x>0,
\end{eqnarray}
where the Riemann-Liouville operator is defined as \cite{Pod} $_{0}D_{t}^{1-\alpha} P(x,t) = \Gamma^{-1}(\alpha) \; \partial/\partial t \int_{0}^{t} dt' P(x,t') \; (t-t')^{\alpha-1}$. The fractional diffusion equation Eq.\ (\ref{anomaleq}) with $\alpha^{-}=\alpha^{+}=\alpha$ and $K^{-}=K^{+}=K$ yields for particles starting on the origin $\left< x^2 \right> = 2 K t^{\alpha}/\Gamma (1+\alpha)$. For $\alpha^{-}=\alpha^{+}=1$ this equation reduces to standard diffusion equation. Without the boundary conditions (soon to be derived),  Eq.\ (\ref{anomaleq}) is nearly useless. The underlying random walk model we consider is the continuous time random walk (CTRW) \cite{BG,MK,CTRW} which is now specified.

{\em Model 2: CTRW.---}Consider a jump process on a discrete lattice with the lattice spacing $a$. For lattice points $x<0$ a particle has the probability $1/2$ to jump to one of its nearest neighbors. Waiting times on each lattice point are independent identically distributed random variables with a common PDF $\psi^{-}(\tau)$. For $x>0$ a similar unbiased random walk takes place with a waiting time PDF $\psi^{+}(\tau)$. On the lattice point $x=0$ (the boundary) a particle has the probability to jump right $q^{+}$ or left $q^{-}=1-q^{+}$ \cite{Cornell} and the waiting times are exponentially distributed with a rate $R_0$. Such biased interface is due for example to a difference of chemical potentials between the two samples \cite{long}. Thus, a particle starting on the origin will jump say to the right (with prob. $q^{+}$) after waiting an average time $1/R_0$, then on the lattice point $x=a$, it will wait for time $\tau$ drawn from $\psi^{+}(\tau)$, and then with probability $1/2$ will jump to the left or right. For subdiffusion the waiting times have power law PDFs $\psi^{-}(\tau) \propto \tau^{-(1+\alpha^{-})}$ and $\psi^{+}(\tau) \propto \tau^{-(1+\alpha^{+})}$, as $\tau \rightarrow \infty$. More specifically, using standard Tauberian theorem the Laplace transform $\tau \rightarrow s$ of the waiting time PDFs behave like $\psi^{-}(s) \sim 1 - B^{-} s^{\alpha^{-}}$, $\psi^{+}(s) \sim 1 - B^{+} s^{\alpha^{+}}$ when $s \rightarrow 0$ corresponding to $\tau \rightarrow \infty$ \cite{MK}. All along this work we denote the Laplace transform by the variable in the parentheses $f(s)=\int_0^{\infty} dt \; e^{-st} f(t)$. The generalized diffusion constants are given by $K^{-} = \lim_{a^2 \rightarrow 0, B^{-} \rightarrow 0} a^2/ 2 B^{-}$ and $K^{+} = \lim_{a^2 \rightarrow 0, B^{+} \rightarrow 0} a^2/2 B^{+}$ \cite{MBK00}. Our results are not changed if on $x=0$ the waiting times are power law distributed like $\psi^{-}$ or $\psi^{+}$ instead of exponential. Large number of applications of the CTRW model are discussed in \cite{BG,MK,CTRW}.

\begin{figure}[t!]
\centerline{\psfig{figure=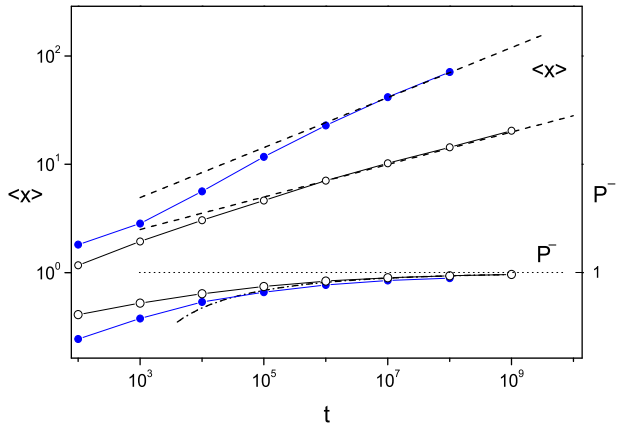,width=90mm,height=70mm}}
\caption{(color online). The drift, $\left< x \right>$, and occupation fraction in $x<0$, $\mathcal{P}^{-}(t)$, calculated numerically for the CTRW model (open circles) with $\alpha^{+}=0.75$, $K^{+}=0.138$, $\alpha^{-}=0.3$, $K^{-}=0.385$, $q^{+}=0.7$ and for the quenched trap model (filled circles) with $\gamma^{+}=0.9$, $\gamma^{-}=0.3$ and $q^{+}=0.7$ (averaged over $20$ realizations of disorder). Dashed lines represent long time asymptotic Eqs.\ (\ref{meanlong}) and (\ref{xqtm}). The dashed-dotted line given by Eq.\ (\ref{p-}) describes how $\mathcal{P}^{-}$ approach its limit $\mathcal{P}^{-} \rightarrow 1$. Notice that all particles flow to the left ($\mathcal{P}^{-} \rightarrow 1$), however $\left< x(t) \right> >0$ namely particles drift to the right.}
\label{FIG1}
\end{figure}
{\em The drift $\left< x \right>$.---}Using the CTRW approach we now calculate the drift. The position of a particle is $x=\sum_{i=0}^{N}\delta x_i$, where $\delta x_i$ is the $i$th displacement and $N$ is the random number of steps. Since the motion is unbiased in domains $x<0$ and $x>0$, we have
\begin{equation}
\label{drift}
\left< x(t) \right> = a (q^{+}-q^{-}) \left< n_z(t) \right>,
\end{equation}
where $\left< n_z(t) \right>$ is the average number of times a particle visited the origin. We define a three state
process $\xi(t) = 0$ if the particle is on the origin, $\xi(t) = +1$ if the particle is in $x > 0$ and $\xi(t) = -1$ if the particle is in $x < 0$. In the long time limit the number of visits to the origin is independent of $R_0$ since
the average waiting times in state $+$ and $-$ are infinite. The waiting times in states $+$ and $-$ are the first passage times \cite{Redner} from $x = a$ to $x = 0$ and from $-a$ to $0$, respectively. These first passage times in the continuum limit are one sided L\'evy distributions whose long time (small $s$) Laplace transforms are \cite{Barkai01} $\phi^{-}(s) \sim 1 - a s^{\alpha^{-}/2}/\sqrt{K^{-}}$ for $x<0$ and similarly $\phi^{+}(s) \sim 1 - a s^{\alpha^{+}/2}/\sqrt{K^{+}}$ for $x>0$. The Laplace transform of the probability to have exactly $n_z$ transitions to state $\xi(t)=0$ is easily found using the Laplace transform convolution theorem \cite{Feller} 
\begin{equation}
\label{P}
P_{n_z}(s) = \frac{1-\bar{\phi}(s)}{s} \bar{\phi}^{n_z}(s), 
\end{equation}
where $\bar{\phi}(s) = q^{-} \phi^{-}(s) + q^{+} \phi^{+}(s)$. From Eq.\ (\ref{P})
\begin{equation}
\label{nav}
\left< n_z (s)\right>  = \frac{\bar{\phi}(s)}{s \left( 1 - \bar{\phi}(s)\right)}. 
\end{equation}
Using the small $s$ expansion of $\left< n_z \right>$ and Eq.\ (\ref{drift}) we obtain for $\alpha^{-}<\alpha^{+}$
\begin{equation}
\label{meanlong}
\left< x (t)\right> \sim \frac{(q^{+} - q^{-})}{q^{-}} \frac{\sqrt{K^{-}}}{\Gamma(1+\alpha^{-}/2)} \; t^{\alpha^{-}/2},
\end{equation}
which agrees well with simulation in Fig.\ \ref{FIG1}. Similar expression is found for $\alpha^{+}<\alpha^{-}$. The sign of the drift, i.e. its directionality, is determined by the sign of $q^{+} - q^{-}$, and $\left< x \right>=0$ if $q^{+}=q^{-}$. Eq.\ (\ref{meanlong}) shows that in the long time limit the drift depends only on one diffusion constant in sample $(-)$ and grows in time with the exponent of the slower medium. This is a surprising result: $\left< x(t)\right>$ can be very far from the interface, deep in the faster sample $x>0$, but still is independent of the properties of that region $\alpha^{+}$, $K^{+}$.

\begin{figure}[t!]
\centerline{\psfig{figure=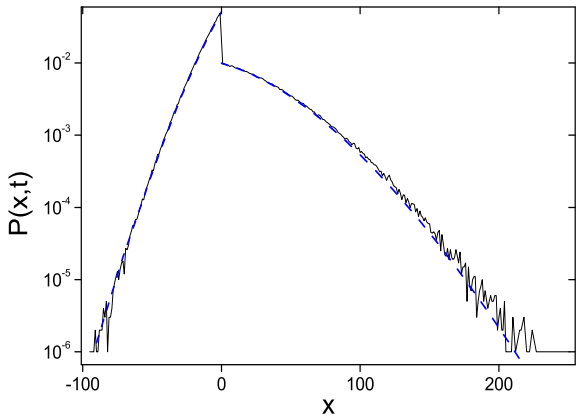,width=90mm,height=70mm}}
\caption{(color online). PDF of particle's position $P(x,t)$ calculated numerically by CTRW model with $\alpha^{-}=0.5$, $K^{-}=0.282$, $\alpha^{+}=0.75$, $K^{+}=0.138$ and $q^{+}=0.7$ at time $t=10^5$. Dashed lines represent solution of fractional diffusion equations Eq.\ (\ref{anomaleq}) calculated by numerical Laplace inversion of Eqs.\ (\ref{anomal2},\ref{eq11}). Notice the jump at the boundary.}
\label{FIG0}
\end{figure}

{\em Boundary Conditions and Solution of Model 1).---}Using the initial condition given by  $P(x,0)=\delta(x)$ the solution of Eq.\ (\ref{anomaleq}) in Laplace space is given by 
\begin{eqnarray}
\label{anomal2}
P(x,s) = C^{+}(s) \frac{s^{\alpha^{+}/2-1} \exp \left( - \frac{|x| s^{\alpha^{+}/2}}{\sqrt{K^{+}}} \right) }{2 \sqrt{K^{+}}} \theta(x) + \nonumber \\
+ C^{-}(s) \frac{s^{\alpha^{-}/2-1} \exp \left( - \frac{|x| s^{\alpha^{-}/2}}{\sqrt{K^{-}}} \right) }{2 \sqrt{K^{-}}} \left[ 1 - \theta(x) \right],
\end{eqnarray}
where $\theta(x)$ is the step function. To find $C^{+}(s)$ and $C^{-}(s)$ we need two boundary conditions. The conservation of probability $\int dx P(x,s)=1/s$ gives
\begin{equation}
\label{eq10}
C^{-}(s) + C^{+}(s) = 2.
\end{equation}
From Eq.\ (\ref{eq10}) we get the first boundary condition which is simply the conservation of the probability current at the boundary 
\begin{equation}
\label{J}
J^{+}(x=0^{+},t) - J^{-}(x=0^{-},t) = \frac{1}{2} \delta(t),
\end{equation}
where $J^{-}(x,t) = - K^{-} \; _{0}D_{t}^{1-\alpha^{-}} \partial P(x,t)/\partial x$ for $x<0$ and similarly for $x>0$ \cite{MBK99}. To derive the second boundary condition we calculate the first moment, $\left< x(s) \right> = \int dx \; x \; P(x,s)$, using Eq.\ (\ref{anomal2})
\begin{equation}
\label{MPDF}
\left< x(s) \right> = \frac{1}{2s} \left( \sqrt{K^{+}} C^{+}(s) s^{-\frac{\alpha^{+}}{2}} - \sqrt{K^{-}} C^{-}(s) s^{-\frac{\alpha^{-}}{2}} \right).
\end{equation}
We require Eq.\ (\ref{MPDF}) to be equal to $\left< x(s) \right>$ Eq.\ (\ref{meanlong}) calculated from the CTRW model. For $\alpha^{-}<\alpha^{+}$, Eqs.\ (\ref{meanlong},\ref{eq10},\ref{MPDF}) yields when $s \rightarrow 0$ 
\begin{equation}
\label{eq11}
C^{+}(s) \sim \frac{2q^{+}}{q^{-}}\sqrt{\frac{K^{-}}{K^{+}}} s^{\frac{\alpha^{+}-\alpha^{-}}{2}}, \; \; C^{-}(s) = 2 - C^{+}(s).
\end{equation}
Inverting Eqs.\ (\ref{anomal2},\ref{eq11}) to the time domain, analytical solutions are in excellent agreement with numerical simulations of the underlying CTRW model (see Fig.\ \ref{FIG0}). For the special case $\alpha^{-}=\alpha^{+}$, we get
\begin{equation}
\label{eq11a}
C^{+}(s) = \frac{2}{1+\frac{q^{-}}{q^{+}} \sqrt{\frac{K^{+}}{K^{-}}}}, \; \; C^{-}(s) = 2 - C^{+}(s).
\end{equation}
Using Eqs.\ (\ref{anomal2},\ref{eq11}) or (\ref{eq11a}), we derive the second boundary condition 
\begin{equation}
\label{eq6}
q^{+} K^{-} s^{-\alpha^{-}} P(x=0^{-},s) = q^{-} K^{+} s^{-\alpha^{+}} P(x=0^{+},s),
\end{equation}
which shows that generally the PDF at the boundary is not continuous, similar to the normal diffusion case \cite{Cornell}. Such behavior is demonstrated in Fig.\ \ref{FIG0}. Note that the boundary condition Eq.\ (\ref{eq6}) has a form of convolution when $\alpha^{+} \ne \alpha^{-}$.

{\em Occupation Fractions.---}From Eqs.\ (\ref{anomal2},\ref{eq11}) occupation fractions, that is probabilities to be in $x<0$ or in $x>0$ are $\mathcal{P}^{-}(s) = \int_{-\infty}^{0} dx \; P(x,s) = C^{-}(s)/(2 s)$ and $\mathcal{P}^{+}(s) = \int_{0}^{\infty} dx \; P(x,s) = C^{+}(s)/(2 s)$. In the long time limit equivalent to $s \rightarrow 0$
\begin{equation}
\label{avfrac}
\mathcal{P}^{+}(s) \sim \frac{1}{s(1+\mathcal{R}^{-1}(s))}, \; \; \; \mathcal{P}^{-}(s) \sim \frac{1}{s(1+\mathcal{R}(s))},
\end{equation}
with $\mathcal{R}(s) = (q^{+} \sqrt{K^{-}})/(q^{-} \sqrt{K^{+}}) \; s^{(\alpha^{+}-\alpha^{-})/2}$.
Let us assume $\alpha^{-}<\alpha^{+}$. Since $\mathcal{P}^{-}(s) \sim 1/s$ when $s \rightarrow 0$, then the probability to be in $x<0$ is
\begin{equation}
\label{p+}
\mathcal{P}^{-}(t) \sim 1, \; \; \; t \rightarrow \infty,
\end{equation}
which indicates that in the long time limit all particles flow to the region $x<0$, where the diffusion is slower (see Fig.\ \ref{FIG1}). Similar result can be obtained from the CTRW model \cite{long}
\begin{equation}
\label{p-}
\mathcal{P}^{-}(t) \sim 1 - \frac{q^{+} \sqrt{K^{-}}}{q^{-} \sqrt{K^{+}}} \frac{t^{\frac{\alpha^{-}-\alpha^{+}}{2}}}{\Gamma \left( 2 + \frac{\alpha^{-}-\alpha^{+}}{2} \right)}.
\end{equation}
Eq.\ (\ref{p-}) is in excellent agreement with numerical simulations (see Fig.\ \ref{FIG1}). 

{\em Paradox and its explanation.---}(i) We observe an averaged net drift from one material to another even if {\em all} particles eventually flow in the opposite direction. If $\alpha^{-}<\alpha^{+}$ and $q^{-}<q^{+}$ the drift $\left< x \right>$ is positive (see Eq.\ (\ref{meanlong}) and Fig.\ \ref{FIG1}). However, the particles are accumulating in $x<0$, that is $\mathcal{P}^{-} \rightarrow 1$ (see Eq.\ (\ref{p+}), Fig.\ \ref{FIG1}). So, we have the drift directed opposite to the flux of the particles {\em even if all particles are eventually in the slower sample.} (ii) If $\alpha^{-}<\alpha^{+}$ but the boundary is unbiased $q^{-}=q^{+}$, in the long time limit the particles will be concentrated in the region $x<0$. However, since $q^{-}=q^{+}$ there is no drift, $\left< x \right> = 0$. An explanation of these paradoxes is as follows: Although the region with smaller $\alpha$ will accumulate more and more particles in the long time limit there will be always some particles in the opposite region where $\alpha$ is larger. These particles are moving more freely and travel far away from the interface which will compensate the accumulation of particles in the region with smaller $\alpha$. In other words while $\mathcal{P}^{+} = 1 - \mathcal{P}^{-} = \int_{0}^{\infty} P(x,t) dx \rightarrow 0$, $\int_{0}^{\infty} x \; P(x,t) dx$ does not approach zero. While dynamics in faster domain ($x>0$) is clearly important (since $\left< x \right> >0$), note that $\left< x \right>$ is independent of the diffusion properties of domain $x>0$. Surprisingly, $\left< x \right>$ Eq.\ (\ref{meanlong}) does not depend on $\alpha^{+}$ and $K^{+}$ as mentioned.


{\em Model 3: Quenched trap model.---}We proceed to show that effects discussed for the CTRW model and fractional diffusion equation are found also for systems with quenched disorder. Consider the quenched trap model where a particle is undergoing a one-dimensional random walk on a quenched random energy landscape on a lattice \cite{BG,qtm}. On each lattice point a random energy $E_x$ is assigned, which is minus the energy of the particle on site $x$, so $E_x>0$ is the depth of a trap on site $x$. The energies of the traps are independent identically distributed random variables with a common PDF $\rho(E)=(1/T_g) \exp(-E/T_g)$. Once the energy at some site $x$ is defined it stays constant in time (quenched disorder), which makes a difference with the corresponding ``annealed" CTRW problem. The lattice is coupled to a heat bath with temperature $T$ which leads to particles escape from site $x$ and jumps to one of its nearest neighbors. The average time it takes the particle to escape from site $x$ is given by Arrhenius law $\tau_x=\exp(E_x/T)$. A small change in $E_x$ leads to exponential change in $\tau_x$. The PDF of the waiting times can be easily calculated $\psi(\tau)= \gamma \tau^{-(1+\gamma)}$ with $\gamma=T/T_g$ and $\tau\ge1$. For $\gamma<1$ ($T<T_g$), all the moments of $\psi(\tau)$ diverge which leads to anomalous diffusion \cite{BG,MK} aging \cite{ag}, and nontrivial occupation times \cite{Burov}. After averaging over different realizations of disorder the evolution of the PDF and the mean squared displacement $\left< x^2 \right>$ of the model is described by the subdiffusive exponent given by $2\gamma/(1+\gamma)$ \cite{BG}. 

We define our composite quenched trap model such that we have $\gamma^{-}$ in $x<0$ and $\gamma^{+}$ in $x>0$ and an interface located at $x=0$. On the boundary a particle has the probability to jump right $q^{+}$ or left $q^{-}=1-q^{+}$ and the waiting times are exponentially distributed. Numerical simulations reveal the behavior that was found for CTRW model and fractional diffusion equations Eq.\ (\ref{anomaleq}). Namely, for $\gamma^{-}<\gamma^{+}$ we find the drift ($\mathcal{C}$ is a constant)
\begin{equation}
\label{xqtm}
\left< x \right> \sim \mathcal{C} (q^{+} - q^{-}) \; t^{\gamma^{-}/(1+\gamma^{-})}, 
\end{equation}
which is positive for $q^{+}>q^{-}$, while almost all particles are found in $x<0$, $\mathcal{P}^{-} \rightarrow 1$ as shown in Fig.\ \ref{FIG1}.
\hfill


{\em Remark.---}Boundary conditions of fractional diffusion equations and CTRW models are non-trivial and have attracted previous interest \cite{Silbey}. We note that the solution of fractional equation Eqs.\ (\ref{anomal2},\ref{eq11}) must be used with care. While this solution gives the correct asymptotic behavior of the occupation fraction $\mathcal{P}^{-} \rightarrow 1$ (when $\alpha^{-}<\alpha^{+}$) (see Eq.\ (\ref{p+})), the correction term within the fractional framework is $\mathcal{P}^{-}(t) \sim 1 - q^{+} \sqrt{K^{-}}/(q^{-} \sqrt{K^{+}}) \; \Gamma^{-1} \left( 1 + \frac{\alpha^{-}-\alpha^{+}}{2} \right) \; t^{\frac{\alpha^{-}-\alpha^{+}}{2}}$. This correction term is different from the exact CTRW result Eq.\ (\ref{p-}) (compare the Gamma functions). Thus, fractional equation works in the long time limit and already leading corrections to asymptotic solution show deviations from exact result.


To summarize, we investigated infiltration in subdiffusive systems. Using the CTRW model we derived the boundary conditions of the problem which allow analytical solution of the fractional diffusion equations. Particles flow to the slower medium while the direction of the averaged drift is determined by breaking of symmetry, $q_L \ne q_R$ in our model. This leads to interesting phenomena unique to anomalous diffusion: (i) all particles are found in one sample ($\mathcal{P}^{-} \rightarrow 1$), but the drift is oppositely directed ($\left< x \right> > 0$), (ii) drift does not depend on properties of fast medium ($\left< x \right>$ is independent of $\alpha^{+}$, $K^{+}$ even though $\left< x \right>$ might be located deep in that medium). We observe similar behavior for the composite quenched trap model which points out to a broader generality of our results. 

This work was supported by the Israel Science Foundation. We thank David~Kessler, Stas~Burov and Shai Carmi for discussions. 

\end{document}